\documentclass[utf8]{frontiersSCNS} 

\usepackage[TU]{fontenc}
\usepackage{url,hyperref,lineno,microtype,subcaption}
\usepackage[onehalfspacing]{setspace}
\usepackage{caption,graphics}
\usepackage{ulem}
\usepackage{soul}
\usepackage{threeparttable}
\usepackage{academicons}
\usepackage{multirow}

\definecolor{orcidlogocol}{HTML}{A6CE39}
\def\keyFont{\fontsize{8}{11}\helveticabold }
\def\firstAuthorLast{Wu {et~al.}} 
\def\Authors{Qi-Qi Wu\,$^{1,2}$, Shi-Long Liao\,$^{1,2,*}$, Xiang Ji\,$^{1,2}$, Zhao-Xiang Qi\,$^{1,2}$, Zhen-Ya Zheng\,$^{1,2}$, Ru-Qiu Lin\,$^{1,2}$,  Ying-Kang Zhang\,$^{1,2}$ and Tao An\,$^{1,2}$}
\def\Address{$^{1}$Shanghai Astronomical Observatory, Chinese Academy of Sciences, Shanghai, P.R.China \\
	$^{2}$School of Astronomy and Space Sciences, University of Chinese Academy of Sciences, Beijing, P.R.China }
\def\corrAuthor{Shi-Long Liao}

\def\corrEmail{shilongliao@shao.ac.cn}

\begin{document}
	\onecolumn
	\firstpage{1}
	
	\title[Abnormal quasars from $Gaia$ EDR3]{SQUAB -- \uppercase\expandafter{\romannumeral1}: The first release of Strange QUasar candidates with ABnormal astrometric characteristics from $Gaia$ EDR3 and SDSS } 
	
	\author[\firstAuthorLast ]{\Authors} 
	\address{} 
	\correspondance{} 
	
	\extraAuth{}

	\maketitle

	\begin{abstract}
		
		\section{}
		
		Given their extremely large distances and small apparent sizes, quasars are generally considered as objects with near-zero parallax and proper motion. However, some special quasars may have abnormal astrometric characteristics, such as quasar pairs, lensed quasars,  AGNs with bright parsec-scale optical jets, which are scientifically interesting objects, such as binary black holes. These quasars may come with astrometric jitter detectable with $Gaia$ data, or significant changes in the position at different wavelengths. In this work, we aim to find these quasar candidates from $Gaia$ EDR3 astrometric data combining with Sloan Digital Sky Survey (SDSS) spectroscopic data to provide a candidate catalog to the science community. We propose a series of criteria for selecting abnormal quasars based on $Gaia$ astrometric data. We obtain two catalogs containing 155 sources and 44 sources, respectively. They are potential candidates of quasar pairs.

		\tiny
		\keyFont{ \section{Keywords:} astrometry, catalogs, quasars, quasar pairs, reference frame} 
	\end{abstract}
	
	\section{Introduction}
	\label{sec:introduction}
	
	Since the discovery of the first quasar in 1963 \citep{schmidt19633}, this type of extremely distant active galactic nuclei (AGN) has gradually become the focus of astronomical research. In astrometry, a large number of evenly distributed quasars can be used to establish a celestial reference frame \citep{ma1997realization,ma2009second,mignard2018gaia,charlot2020third} because they have almost zero proper motions and point-like shapes. On the other hand, quasars are also a critical pathway to explore the evolution and mergers of galaxies in astrophysics \citep{begelman1980massive,shen2021hidden}.
	
	There are many surveys concerning the identification of quasars such as the large Bright Quasar Survey \citep{hewett1995large}, the 2DF Quasar Redshift Survey (2QZ, \citealt{croom20042df}), the quasars from Large Sky Area Multi-Object Fiber Spectroscopic Telescope (LAMOST, \citealt{luo2012data}) and Solan Digital Sky Survey (SDSS, \citealt{paris2018sloan,lyke2020sloan}). A large number of quasars have also been identified through astrometry and mid-infrared methods (see, e.g. \citealt{secrest2015identification, guo2018identifying}). The total number of identified quasars has exceeded one million, and these quasars have been collected and compiled into various catalogs (see, e.g. \citealt{veron2010catalogue, souchay2019lqac, liao2019compilation, flesch2021million}). Among these confirmed quasars, some spectroscopically identified quasars show abnormal astrometric characteristics in the $Gaia$  high-precision astrometric observation \citep{wu2021catalog}. These abnormal quasars have large proper motions or significant astrometric noises, which means that they are not suitable to be used to establish the celestial reference frame. \citet{shen2019varstrometry} emphasize that quasars with significant astrometric noises may be dual quasars. These dual quasars are precursors of the binary supermassive black holes, which play an important role in the study of galaxy evolution and gravitational waves (GWs). At present, most of the known dual AGN are at low redshifts or have large physical separation ($>$20kpc), and only several known small-separation dual quasars are at high redshifts \citep{Chen_2022}, while $Gaia$'s high-precision astrometric data has not been seriously considered.
	
	$Gaia$ is an astrometric satellite launched by the European Space Agency (ESA) on 19 December 2013 \citep{prusti2016gaia}. At present, $Gaia$ has provided high-precision astrometric data for more than 1.8 billion sources in the G magnitude range from 3 to 21 mag \citep{lindegren2021gaia}. With the accurate position data and a large number of identified quasars, $Gaia$ has been committed to establishing its own optical non-rotating celestial reference frame (CRF) \citep{mignard2018gaia}. \citet{lindegren2018gaia} selected 556,869 quasars from the 3rd International Celestial Sphere Reference Frame (ICRF3) and AllWISE AGN catalog \citep{secrest2015identification} to establish the $Gaia$-CRF2 (see also \citealt{mignard2018gaia}). In $Gaia$ Early Data Release 3 (EDR3), the AGN catalog, which contains 1,614,173 sources, is obtained by cross-matching with 17 external AGN catalogs. The systematic errors in EDR3 have been greatly improved compared with DR2.  The astrometric properties of the EDR3 quasars  show that no significant residuals are found globally \citep{liao2021probing, liao2021probing1}, which provides us with a unique opportunity to select abnormal quasars in EDR3.

	In EDR3, there are 585 million 5-parameter\footnote{In $Gaia$ EDR3, there are three types of sources according to the astrometric solutions, the position (right ascension and declination), parallax and two components of proper motion are available for 5-parameter sources, the astrometrically estimated effective wavenumber together with the above five parameters are available for 6-parameter sources, and for 2-parameter sources, only positional data are provided \citep{lindegren2021gaia}.} and 882 million 6-parameter sources with the measurement of parallax and proper motion, while the remaining 344 million 2-parameter sources have only positional data. The quasars used by $Gaia$ were obtained by a cross-match of the full $Gaia$ catalog with the external QSO/AGN catalogs,  the matched sources were further selected to have parallaxes and proper motions compatible with zero within five times the respective uncertainty \citep{lindegren2018gaia, klioner2021gaia}. Therefore, among the common sources of $Gaia$ EDR3 and the 14th data release of SDSS Quasars (SDSS DR14Q, \citealt{paris2018sloan}), 308,601 of 367,516 quasars are contained in the $Gaia$ EDR3 AGN catalog. For the remaining 58915 quasars, 206 sources are ruled out due to excessive proper motion or parallax, and 58707 quasars are excluded just because they do not have the measurement of proper motion and parallax. To make full use of the position information of these 2-parameter quasars, we need to judge the reliability of their astronomical information through other criteria.
	
	In this paper, we try to explore the selection of quasars with abnormal astrometric characteristics using different combinations of appropriate astrometric parameters in addition to parallaxes and proper motions. In this way, we can not only evaluate the 5-parameter or 6-parameter sources more comprehensively but also appropriately select the 2-parameter sources to further expand the sample of quasars we can use in $Gaia$. Note that we are not selecting quasars with good observation parameters. On the contrary, we want to mark the quasars with poor astrometric parameters, which will provide some candidates for studying galaxy evolution and binary black holes. 
	
	This paper is organized as follows. In section \ref{sec:Data and Selection}, we introduce the data and criteria for selecting quasars with abnormal astrometric characteristics. We show the results and evaluate these quasars in section \ref{sec:result}. In section \ref{sec:discussion}, we make some discussions about the extension of the catalogs and the identification of quasar pairs, and the conclusions are given in section \ref{sec:conclusion}.

	\section{Data and Selection}
	\label{sec:Data and Selection}
	
	\subsection{Data used}
	
As addressed in the previous section, the AGN catalog in $Gaia$ EDR3 (GEAC hereafter) is obtained by cross-matching with 17 external AGN catalogs.  GEAC contains 1,215,942 5-parameter sources and 398,231 6-parameter sources.  Besides, to calculate the rotation of the $Gaia$ reference frame, the $Gaia$ team selected 429,249 5-parameter solution quasars as frame rotator sources (FRS hereafter, \citealt{brown2021gaia}).  Therefore, FRS is currently the most reliable quasar catalog in $Gaia$, and will be used as a comparison sample to evaluate the astrometric parameters of other quasar candidates.

As mentioned in the previous section, there have been many compiled quasar catalogs. The spectra classified quasars from SDSS contributed a large proportion. Considering the reliability and the indispensable images and spectra data of SDSS, we decided to use the SDSS quasar catalog as our input catalog to select the abnormal quasars. SDSS Data Release 16 (DR16, \citealt{jonsson2020apogee}) is the latest data product from Apache Point Observatory Galactic Evolution Experiment (APOGEE)-2/Sloan Digital Sky Survey-\uppercase\expandafter{\romannumeral4} \citep{blanton2017sloan}. And the quasar catalog of SDSS DR16 \citep{lyke2020sloan} contains two catalogs: the quasar-only catalog and the ``superset'' objects targeted as quasars. The ``superset'' of all SDSS-IV objects targeted as quasars containing 1,440,615 sources and the quasar-only catalog containing 750,414 quasars. Due to the high completeness (99.8\%) and low contamination (0.3\%-1.3\%), we choose the quasar-only catalog as our initial sample of quasars (SDSS DR16Q hereafter).
	
	\subsection{The selection criteria}
	
	With a large number of quasars identified by SDSS spectrum, after cross-match with $Gaia$ EDR3 in a $1^{\prime \prime}$ radius, we obtain 489,402 common sources in $Gaia$ EDR3 and SDSS DR16Q. Among them, there are 153 SDSS quasars with two $Gaia$ matches, two SDSS quasars with three $Gaia$ matches and one $Gaia$ source with two SDSS quasars matches. We then exclude two SDSS quasars whose corresponding four $Gaia$ matched sources are all with significant proper motion or parallax. These sources are compiled into the type A catalog of quasars with abnormal astrometric characteristics (Catalog A hereafter). These multiple-matched sources are potential quasar pairs or star-quasar pairs, which will be further discussed later in this paper.
	
	For the remaining 489,285 quasars with only one $Gaia$ source matched within a $1^{\prime \prime}$ radius, to lower the possibility of star contamination in cross-matching, we exclude those sources with significant proper motion or parallax using the criteria mentioned in the previous section. Then we select several astrometric parameters emphasized in \citet{lindegren2021gaia} to evaluate their accuracy and reliability. These parameters can characterize the goodness of the point spread function (PSF) model fitting of each source and the reliability of the observation data. We will introduce them and describe in detail the criteria of our selection in the following parts.
	
\begin{figure}[h!]
		\begin{center}
			\includegraphics[width=0.7\textwidth]{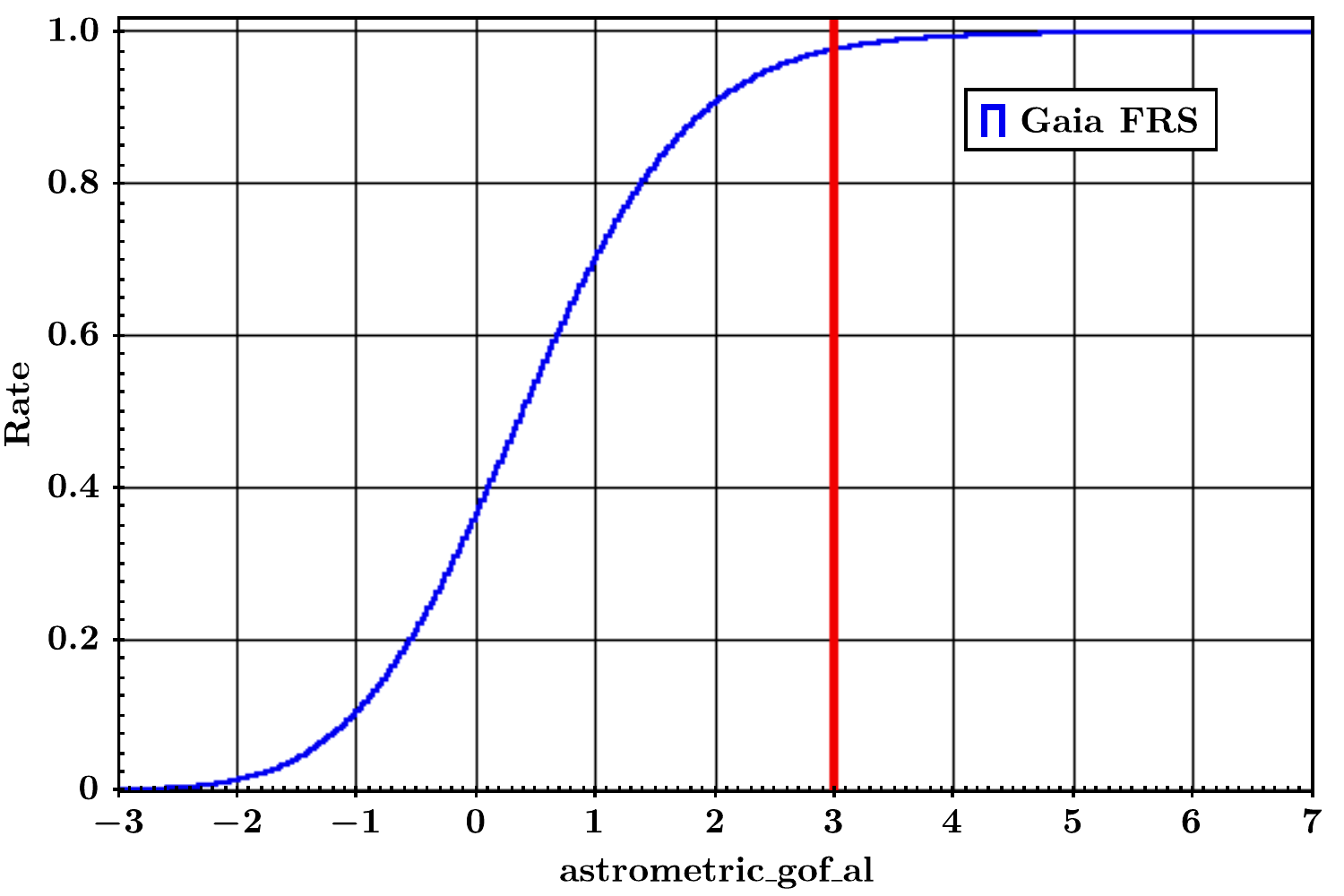}
		\end{center}
		\caption{The cumulative distribution histogram for $astrometric\_gof\_al$ values of sources in $Gaia$ FRS. The red vertical line is $astrometric\_gof\_al$ = 3.}
		\label{fig:gofal}
	\end{figure}

\begin{figure}[h!]
		\begin{center}
			\includegraphics[width=0.7\textwidth]{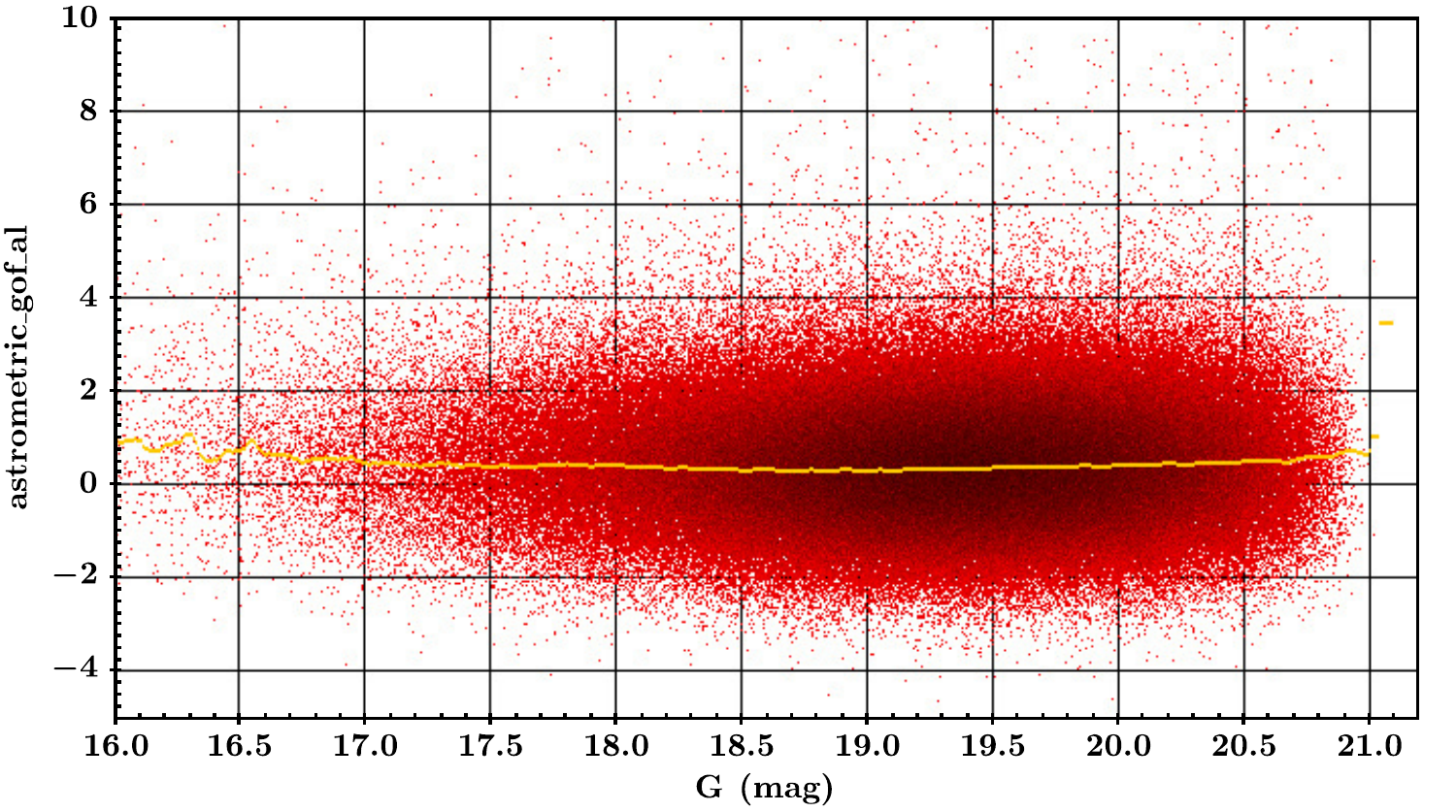}
		\end{center}
		\caption{The distribution of $astrometric\_gof\_al$ with the increase of Gmag. The red dots are data points, and the yellow line is the median line of each bin.}
		\label{fig:gofalmag}
	\end{figure}

$astrometric\_gof\_al$ represents the Goodness-of-fit statistic of the astrometric solution for the source in the along-scan direction.
 \href{https://gea.esac.esa.int/archive/documentation/GEDR3/Gaia_archive/chap_datamodel/sec_dm_main_tables/ssec_dm_gaia_source.html}{The $Gaia$ EDR3 documentation} proposed a rough value of this criterion to distinguish between good and bad fitting of the data: if the $astrometric\_gof\_al$ is greater than 3, it may indicate a bad fitting of the data. We have analyzed the reliability of this criterion by checking the statistical value of $astrometric\_gof\_al$ from FRS sources. There are only about 3\% of quasars in FRS that have the excessive $astrometric\_gof\_al$ ( $>$ 3) as indicated from Fig. \ref{fig:gofal}, which means this criterion could select some extreme quasars while ensuring that most reliable quasars are ruled out. Fig. \ref{fig:gofalmag} shows that the median line of $astrometric\_gof\_al$ is almost parallel to the x-axis, so there is no obvious correlation between $astrometric\_gof\_al$ and the brightness of source when G $<$ 20.9 mag. With these studies in mind, we choose $astrometric\_gof\_al > 3$ as one of the criteria to select the quasars with abnormal astrometric characteristics.

\begin{figure}[h!]
		\begin{center}
			\includegraphics[width=0.7\textwidth]{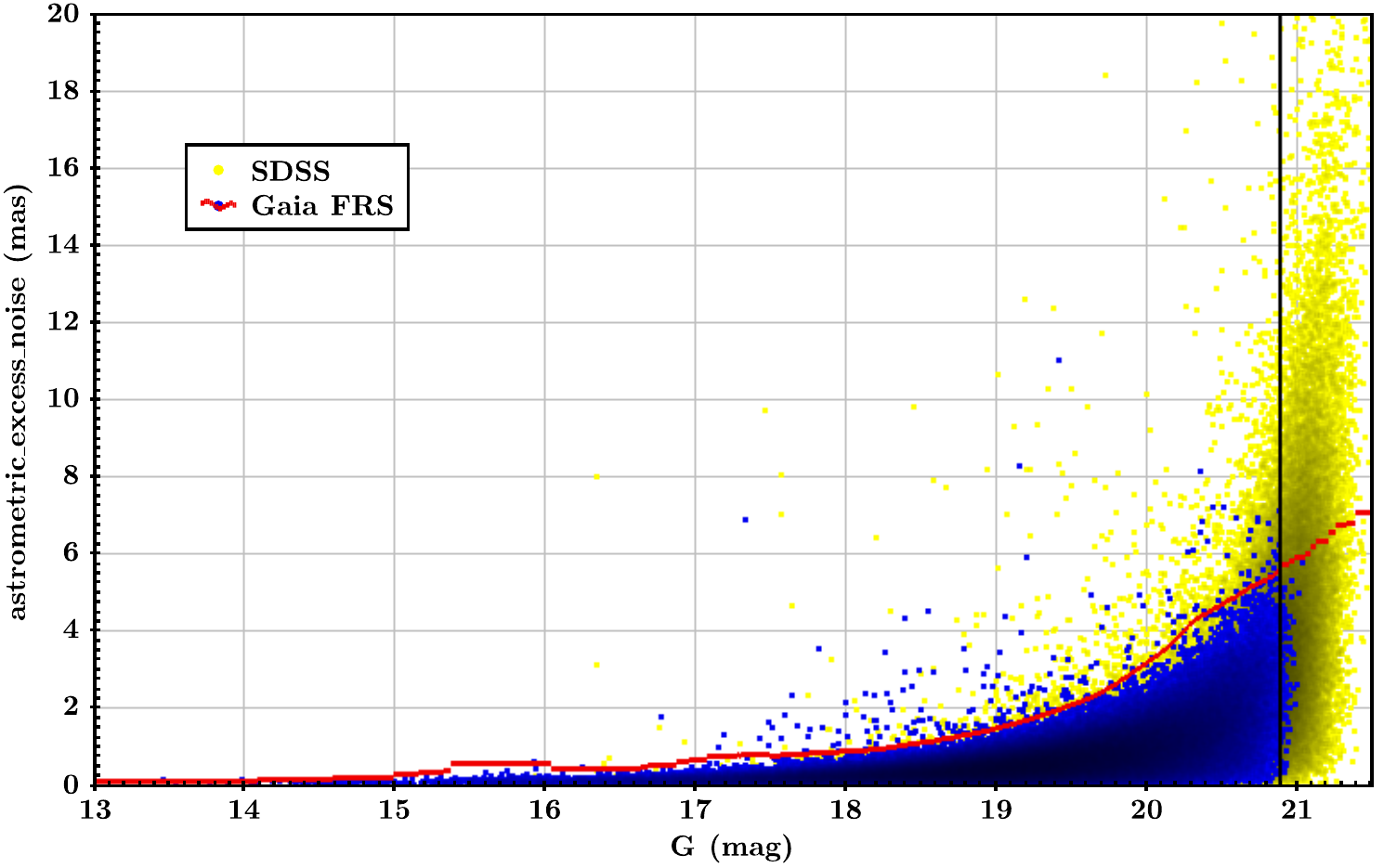}
		\end{center}
		\caption{The $astrometric\_excess\_noise$  versus Gmag. The blue dots represent the sources from $Gaia$ FRS, while the yellow dots are the common sources from $Gaia$ EDR3 and SDSS DR16Q. The red curve is the 99.9\% quantile line of $Gaia$ FRS, and the black vertical line represents G = 20.9 mag.}
		\label{fig:noise}
	\end{figure}

	$astrometric\_excess\_noise$ represents the disagreement between the $Gaia$ observations of a source and the best-fitting standard astrometric model, and a large value signifies that the residuals are statistically larger than expected. There is no doubt  that $astrometric\_excess\_noise$ is an important indicator of whether the source is astrometrically ``well-behaved'', but we need to make sensible cutoffs to ensure that the sources we selected are reliable and logical. With high accuracy and reliability, FRS is an ideal reference to determine the criterion of noise. As seen in Fig. \ref{fig:noise}, with the magnitudes of the sources becoming fainter, the observation noises of the sources are also rapidly increasing. The 99.9\% quantile line can retain most of the reliable quasars, and the blue points outside this line show obvious bias from the whole sample. Therefore, the red curve may be an empirically feasible criterion. We choose the 20.9 mag as the magnitude limit of this criterion since there are only 138 FRS sources fainter than this limit. We plot the quasars of SDSS DR16Q in the same figure and find 1982 of them meet this 99.9\% quantile criterion\footnote{The criterion select the bright sources ($<$20.9 mag) above the 99.9\% quantile line.}. Another parameter that could be used to evaluate the astrometric noise is $astrometric\_excess\_noise\_sig$, which represents the significance of excess noise. 
	Since the excess noise could absorb all kinds of modeling errors such as PSF (Point spread function) calibration errors and geometric instrument calibration errors \citep{lindegren2012astrometric}, the $astrometric\_excess\_noise\_sig$ is important to evaluate if the noise is caused by the structure of the source. The $Gaia$ document recommends that $astrometric\_excess\_noise\_sig > 2$ indicates that the given noise is probably significant. We have not found any obvious correlation between the significance and magnitude in FRS, so $astrometric\_excess\_noise\_sig > 2$ could be the sensible cutoff to ensure the excess noise is applicable for all magnitudes.

\begin{figure}[h!]
		\begin{center}
			\includegraphics[width=0.7\textwidth]{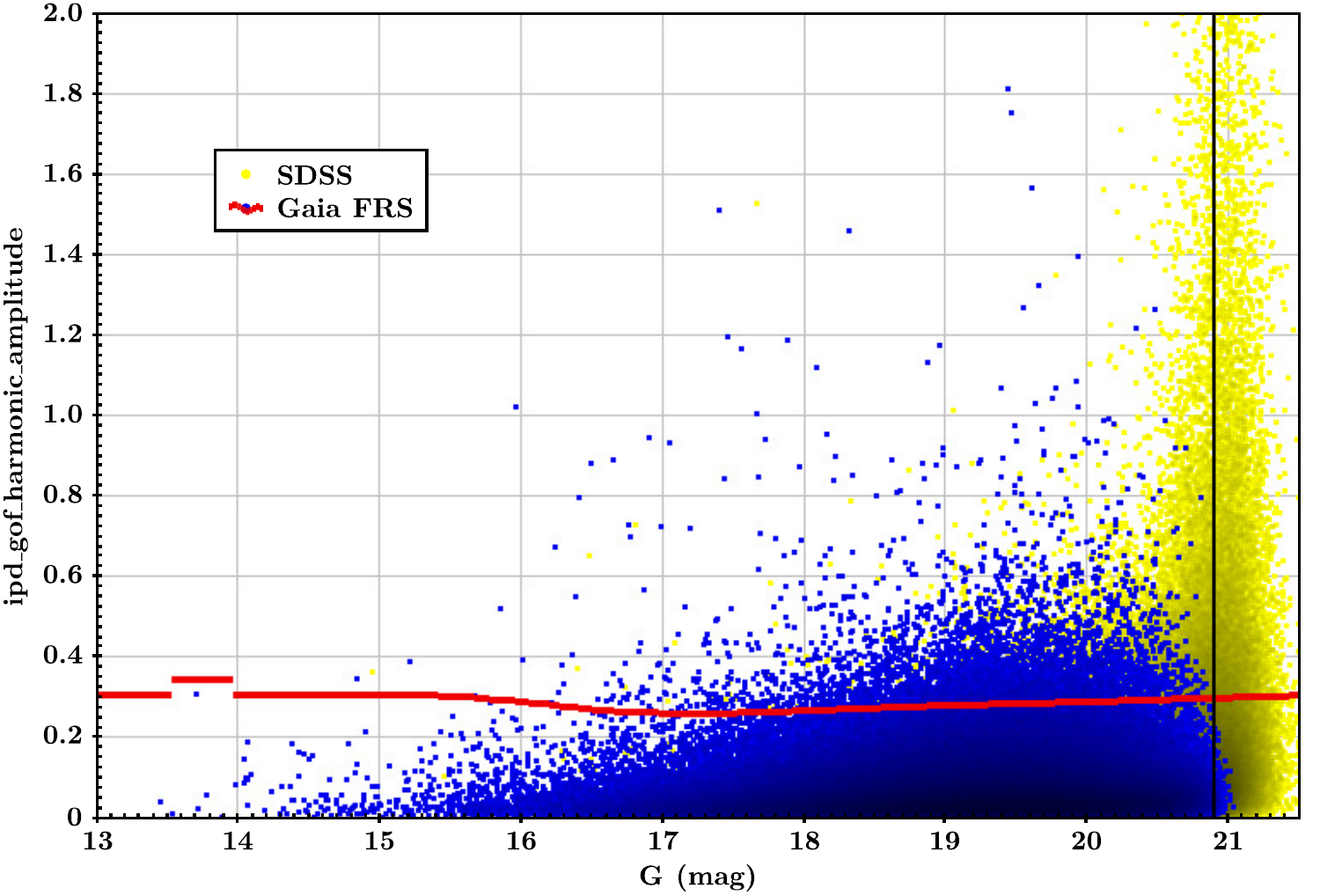}
		\end{center}
		\caption{The $ipd\_gof\_harmonic\_amplitude$  versus Gmag. The blue dots represent the sources from $Gaia$ FRS, while the yellow dots are the common sources from $Gaia$ EDR3 and SDSS DR16Q. The red curve is the 99\% quantile of $Gaia$ FRS, and the black  vertical line represents G = 20.9 mag.}
		\label{fig:amplitude}
	\end{figure}

	$ipd\_gof\_harmonic\_amplitude$ measures the amplitude of the variation of the goodness-of-fit of image parameter as a function of the position angle of the scan direction. A large amplitude might indicate the source has more than one optical center. Quasar pairs, or AGN with bright parsec-scale optical jets, may lead to a relatively large amplitude of the sources, and the positioning accuracy of these quasars could be affected by the multiple centers. We hope to use the same method as for the excess noise to obtain a suitable criterion. As seen in Fig. \ref{fig:amplitude}, it seems that $ipd\_gof\_harmonic\_amplitude$ does not correlate with magnitude, and the 99\% quantile line is almost a straight line parallel to the x-axis. The criterion we selected for this parameter is $ipd\_gof\_harmonic\_amplitude > 0.26$ when G $<$ 20.9 mag.

	$ipd\_frac\_multi\_peak$ is another important parameter for evaluating whether the source is a binary. It provides the percent of successful-IPD (Image Parameters Determination) windows with more than one peak, and we could preliminarily judge whether a source is a visually resolved double star based on this parameter. Normally, all sources with percent greater than zero should be selected as abnormal quasar candidates, and totally, we found that there are 32578 sources in FRS whose $ipd\_frac\_multi\_peak$ is greater than zero, with only 3215 (10\%) of them greater than one. A large number of sources with $ipd\_frac\_multi\_peak = 1$ may increase the contamination of our final catalog, and $ipd\_frac\_multi\_peak > 1$ can be used to select some extreme quasars efficiently. So we take $ipd\_frac\_multi\_peak > 1$ as the criterion: in this case, 3392 (0.7\% of SDSS quasars) quasars are selected.

	With the considerations above, we propose the following criteria for selecting abnormal quasars in SDSS DR16Q:
	
	\begin{equation}
		\left\{
		\begin{array}{lr}
			(\text{i})\quad  astrometric\_gof\_al > 3, &  \\
			(\text{ii})\quad  astrometric\_excess\_noise > \text{$99.9\%$ quantile line of $Gaia$ FRS}, &  \\
			(\text{iii})\quad  astrometric\_excess\_noise\_sig > 2, &  \\
			(\text{iv})\quad  ipd\_gof\_harmonic\_amplitude > 0.26, &  \\
			(\text{v})\quad  ipd\_frac\_multi\_peak > 1, &  \\
			(\text{vi})\quad  G < 20.9\text{ mag}, &  \\
		\end{array}
		\right.\label{criteria}
	\end{equation}
	we finally obtained 44 quasars that met all of the above criteria, and these quasars are included in the type B catalog of abnormal quasars (Catalog B hereafter).
	
	\section{Result}
	\label{sec:result}
\begin{figure}[h!]
	\begin{center}
			\includegraphics[width=0.7\textwidth]{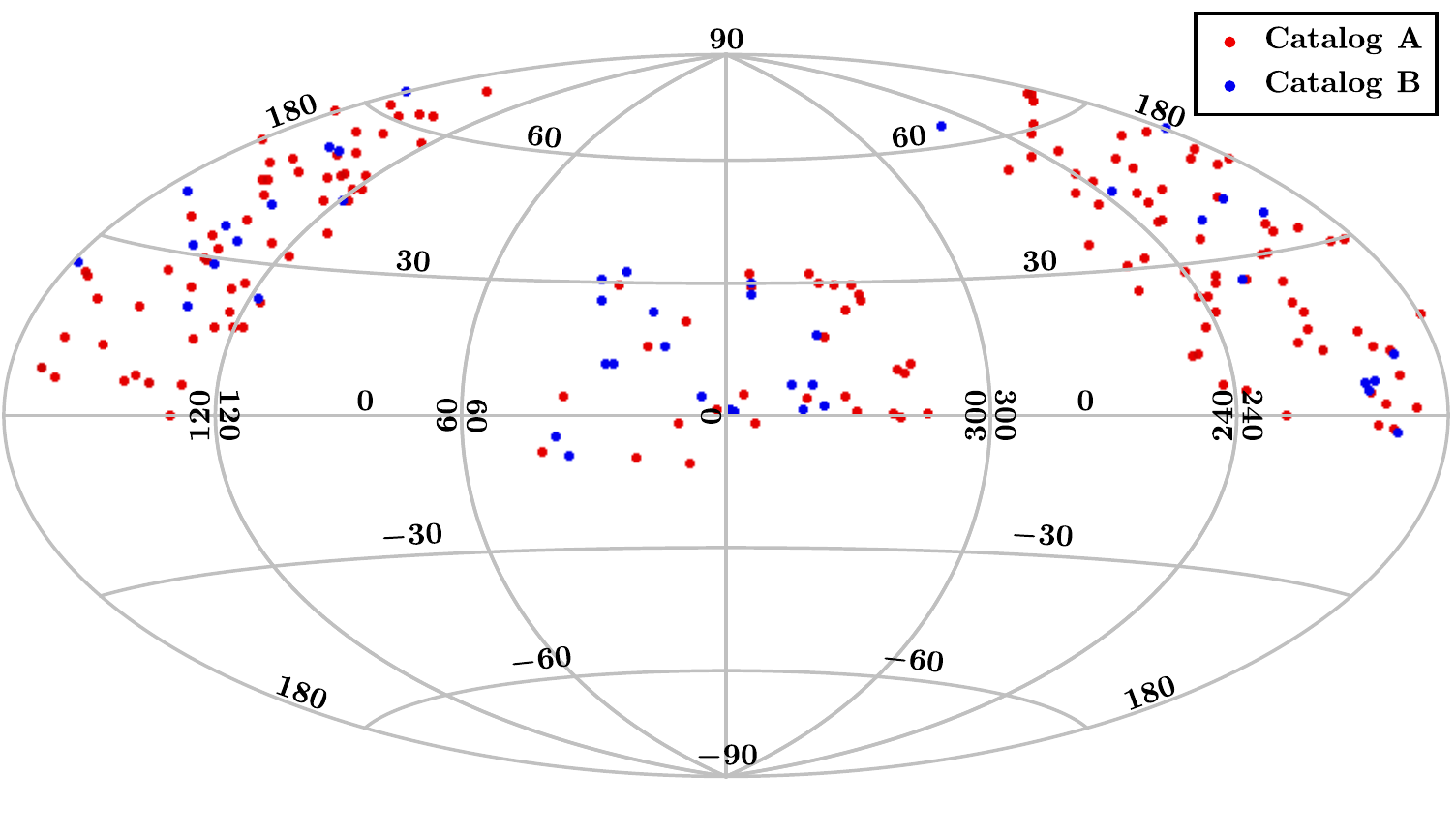}
	\end{center}
		\caption{The sky distribution of the sources in Catalog A and B. The map uses the Hammer Aitoff projection in Equatorial coordinates.}
		\label{fig:distribution}
	\end{figure}

\begin{figure}[!htbp]
	\centering
	\begin{tabular}{c}
		\includegraphics[width=0.43\textwidth]{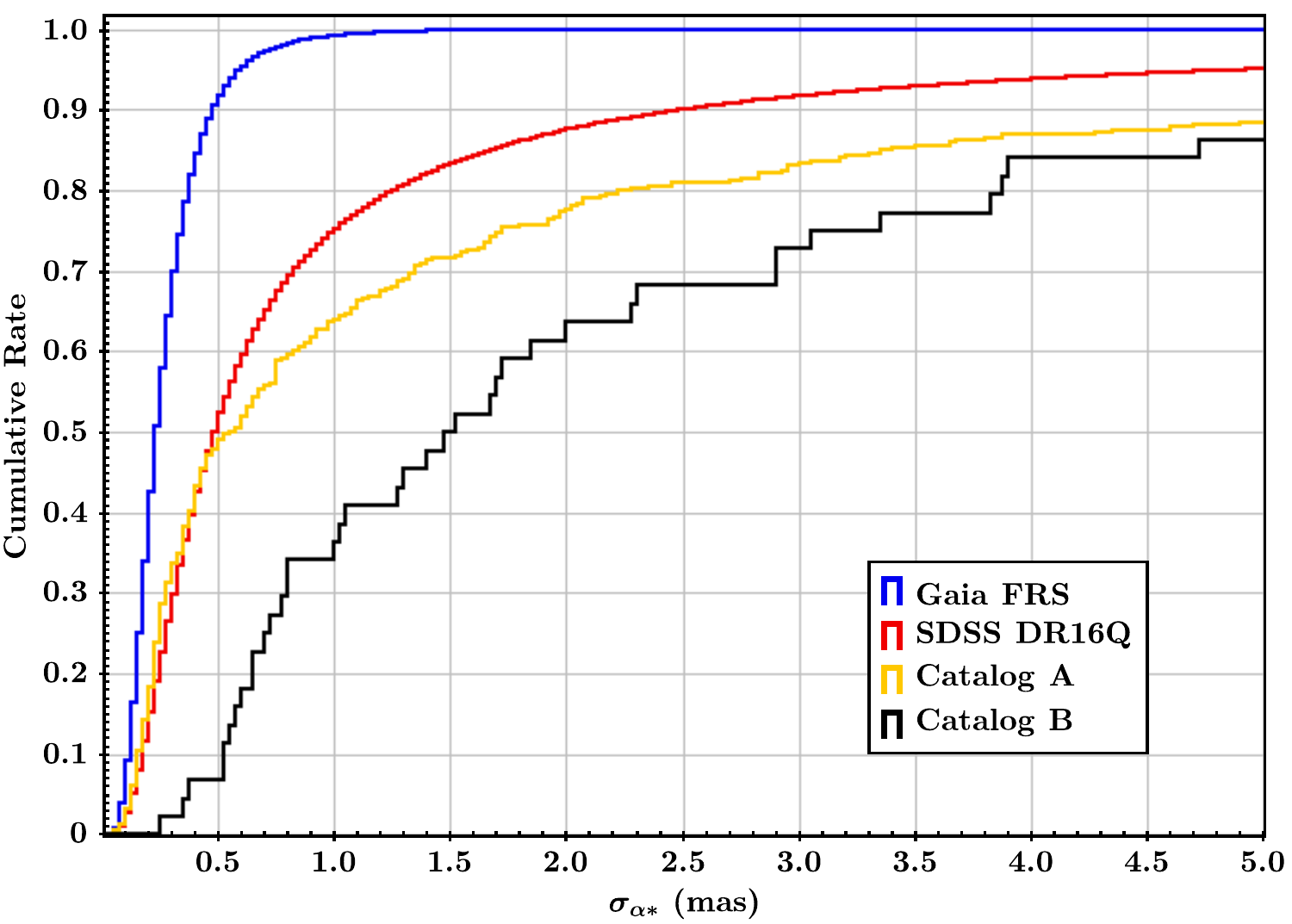}
		\small(A)
	\end{tabular}
	\begin{tabular}{c}
		\includegraphics[width=0.43\textwidth]{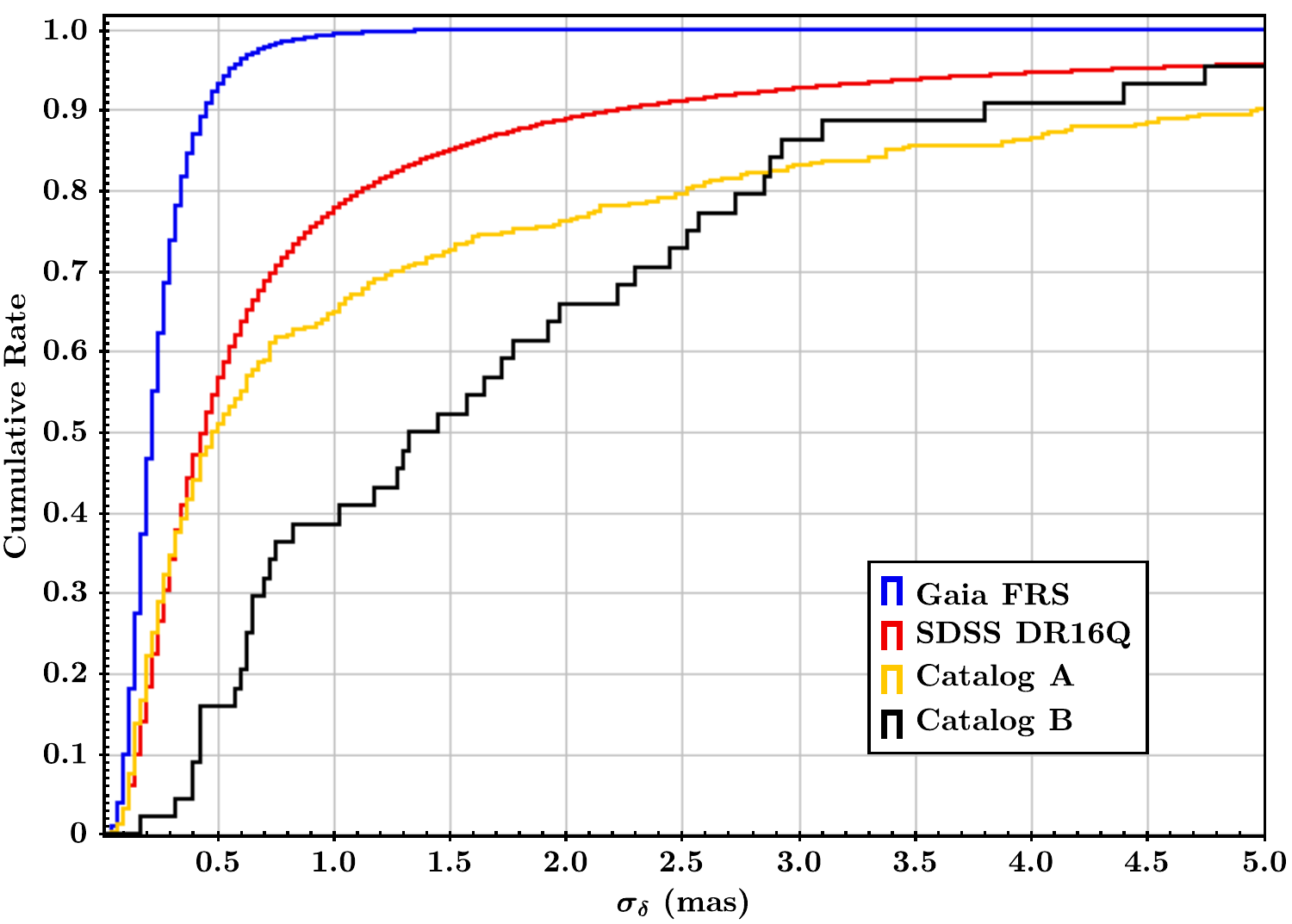}
		\small(B)
	\end{tabular}
	\caption
	{\small The cumulative distribution histogram for $\sigma_{\alpha*}$ (A) and $\sigma_{\delta}$ (B) of sources in $Gaia$ FRS, SDSS DR16Q, Catalog A and Catalog B.}
	\label{fig:positione}%
\end{figure}

\begin{figure}[!htbp]
	\centering
	\begin{tabular}{c}
		\includegraphics[width=0.43\textwidth]{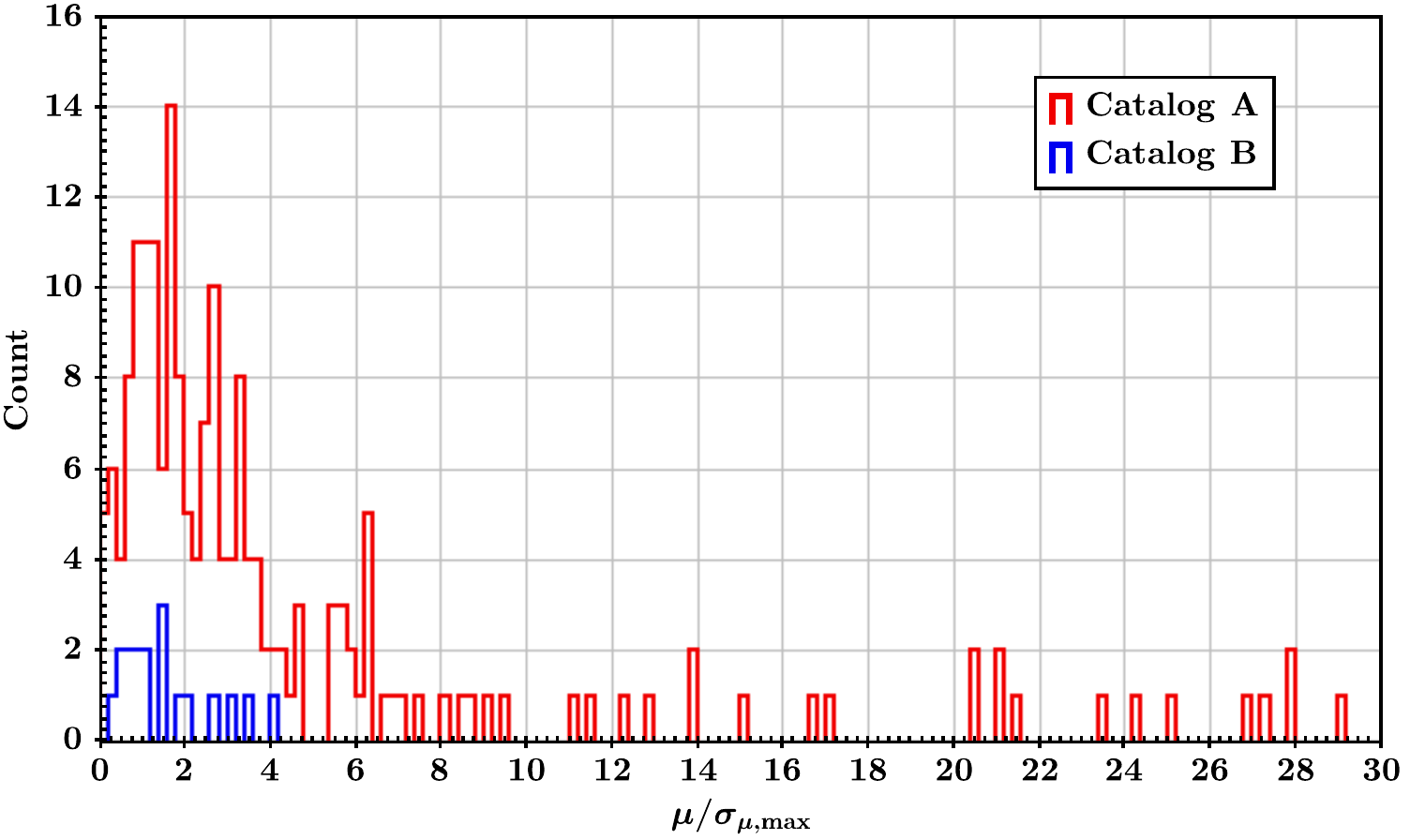}
		\small(A)
	\end{tabular}
	\begin{tabular}{c}
		\includegraphics[width=0.43\textwidth]{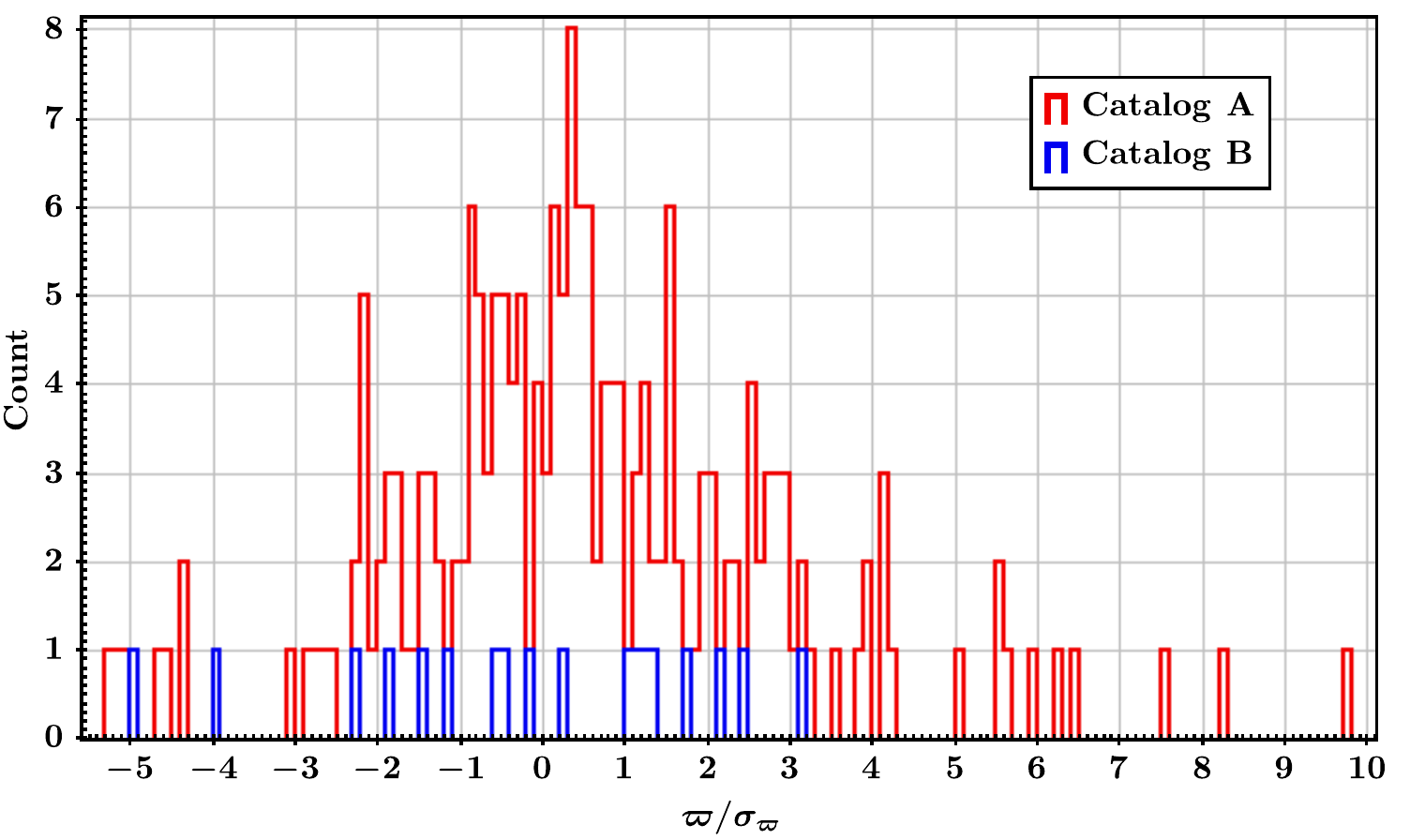}
		\small(B)
	\end{tabular}
	\caption
	{\small The normalized proper motion (A) and parallax $\varpi/\sigma_{\varpi}$ (B) distributions for sources in Catalog A and Catalog B.}
	\label{fig:parapm}%
\end{figure}

\begin{figure}[h!]
		\begin{center}
			\includegraphics[width=0.7\textwidth]{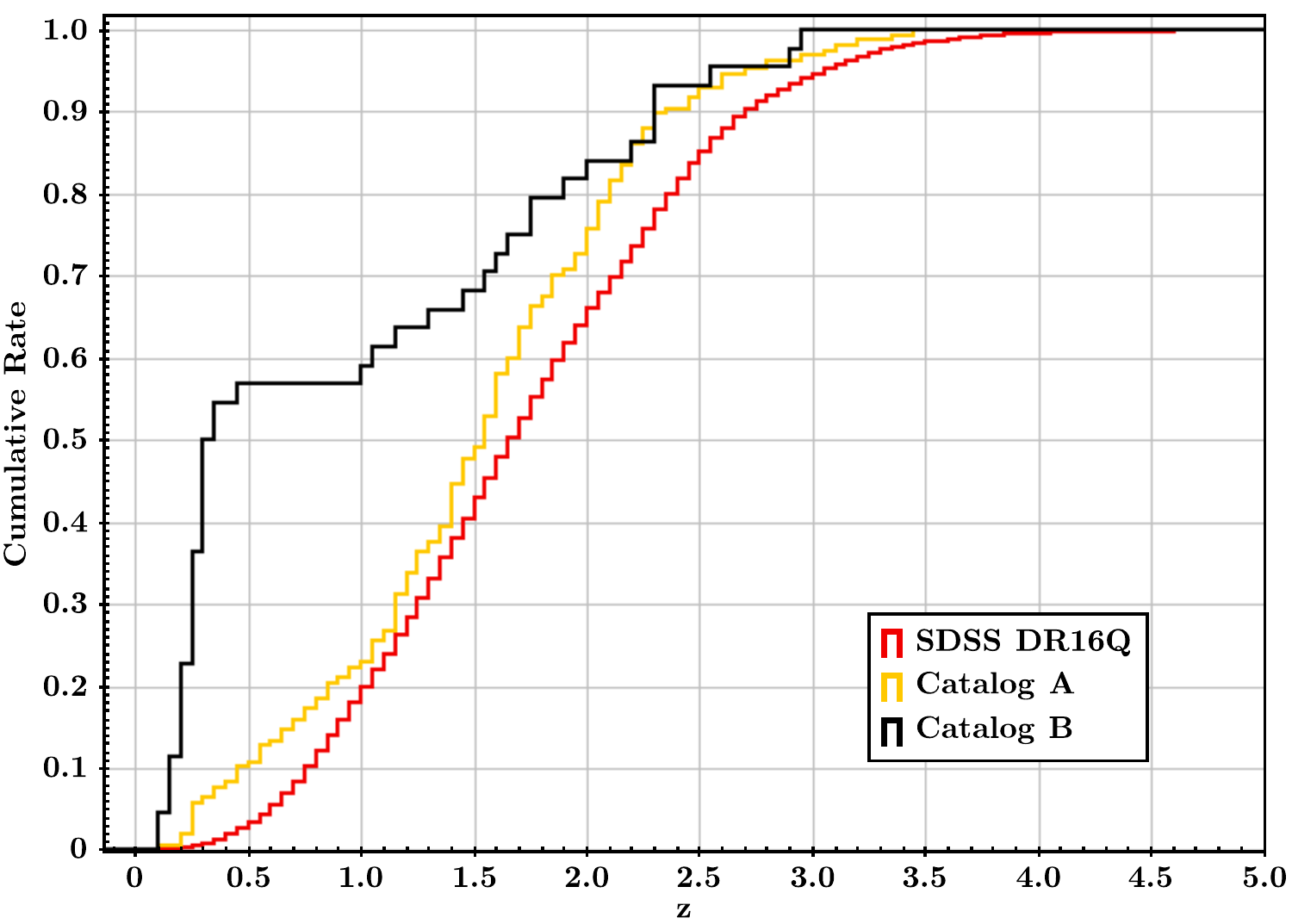}
		\end{center}
		\caption{The cumulative distribution histogram for redshift of sources in SDSS DR16Q, Catalog A and Catalog B.}
		\label{fig:z}
	\end{figure}

	In Table \ref{tab:a} and Table \ref{tab:b} we detail the contents of our catalogs. The sky distribution of the two catalogs is shown in Fig. \ref{fig:distribution}. There are 108/309\footnote{Due to the multiple matches, we obtained 151*2+2*3+1=309 Gaia sources.} (35.0\%) 2-parameter $Gaia$ sources in Catalog A, and for Catalog B, the rate is 26/44 (59.1\%). Therefore, for the two whole catalogs, the position errors are obviously greater than those of $Gaia$ FRS and SDSS DR16Q as expected, see Fig. \ref{fig:positione}. For the 5-parameter and 6-parameter sources in Catalog A and B, the normalized proper motion and parallax distributions are shown in Fig. \ref{fig:parapm}. Compared to the almost zero parallax and proper motion of $Gaia$ FRS, the sources in Catalog A and B have worse astrometric solutions. The $Gaia$ celestial reference frame ($Gaia$-CRF3) is materialised by 1,614,173 quasars in GEAC \citep{brown2021gaia}, and we find that there are 111 common sources between GEAC and catalog A, and 16 common sources with Catalog B, which we recommend removing from GEAC. Fig. \ref{fig:z} shows the redshift distribution of these two catalogs: we find that the distribution of Catalog A and SDSS DR16Q is almost consistent. However, the sources in Catalog B are distributed more in the low redshift part, and almost no sources in Catalog B have a redshift in the range of 0.5-0.8.
	
	As we mentioned above, the spectroscopically identified SDSS DR16Q has a contamination of 0.3\%-1.3\%, which is estimated by implementing the visual inspection of the spectra of a randomly chosen sample \citep{lyke2020sloan}. In Catalog A, for the 155 SDSS spectroscopically identified quasars, 43 have been visually inspected, and 36 are Quasars, while 7 of them are identified as BAL Quasars. Of the remaining 112 sources with only spectral identification, 98 have been included in LQAC5\footnote{The quasars in LQAC5 are compiled from SDSS DR14Q and other quasar catalogs, the newly identified quasars in SDSS DR16Q are not included.}\citep{souchay2019lqac}, and the remaining 14 quasars are newly identified by SDSS DR16Q. In Catalog B, 10 of the 44 SDSS quasars have been visually inspected, and all of them are Quasars.  For the remaining 34 sources with only spectral identification, 24 have been included in LQAC5, and the remaining 10 quasars are newly identified by SDSS DR16Q. Therefore, we believe the quasars in our catalog are reliable.
	
\begin{table}[htbp]
  \centering
  \caption{Description of Catalog A.}
    \label{tab:a}
  \begin{threeparttable}
    \begin{tabular}{|c|c|c|c|}
    \hline
    Lable & Type  & Units & Detail \\
    \hline
    source\_id & long  & -   & Unique source identifier in $Gaia$ EDR3 \\
    \hline
    SDSS  & char  & -   & Unique source identifier in SDSS \\
    \hline
    ra    & double & degree & Right Ascension in J2016.0 \\
    \hline
    dec   & double & degree & Declination in J2016.0 \\
    \hline
    ra\_error & double & mas   & Error of right ascension \\
    \hline
    dec\_error & double & mas   & Error of declination \\
    \hline
    ra\_J2000 & double & degree & Right Ascension of SDSS source in J2000.0 \\
    \hline
    dec\_J2000 & double & degree & Declination of SDSS source in J2000.0 \\
    \hline
    z     & float & -   & Redshift of the matched SDSS source \\
    \hline
    ang   & float & mas   & the angular distance of the two matched sources \\
    \hline
    sign\tnote{*}  & int   & -   & 1, 2, 3 for star-quasar pair, quasar pair and lensing object, respectively \\
    \hline
    \end{tabular}%
  \begin{tablenotes}
  	\footnotesize
  	\item[*] The sign only represents the preliminary classification, not the final identification result. More details about the sign can be found in section \ref{subsec:cataloga}.
  \end{tablenotes}
\end{threeparttable}
\end{table}%

\begin{table}[htbp]
  \centering
  \caption{Description of Catalog B.}
  
    \begin{tabular}{|c|c|c|c|}
    \hline
    Lable & Type  & Units & Detail \\
    \hline
    source\_id & long  & -   & Unique source identifier in $Gaia$ EDR3 \\
    \hline
    SDSS  & char  & -   & Unique source identifier in SDSS \\
    \hline
    ra    & double & degree & Right Ascension in J2016.0 \\
    \hline
    dec   & double & degree & Declination in J2016.0 \\
    \hline
    ra\_error & double & mas   & Error of right ascension \\
    \hline
    dec\_error & double & mas   & Error of declination \\
    \hline
    g\_mag & float & mag   & G-band mean magnitude \\
    \hline
    gof\_al & float & -   & Goodness of fit statistic of model wrt along-scan observations \\
    \hline
    noise & float & mas   & Excess noise of the source \\
    \hline
    noise\_sig & float & -  & Significance of excess noise \\
    \hline
    amplitude & float & -   & Amplitude of the IPD GoF versus position angle of scan \\
    \hline
    multi\_peak & byte  & -   & Percent of successful-IPD windows with more than one peak \\
    \hline
    ruwe  & float & -   & Renormalized unit weight error  \\
    \hline
    duplicated\_source & boolean & -  & Source with multiple source identifiers \\
    \hline
    params\_solved & byte  & -   & 3, 31, 95 for two, five, six parameter sources,  respectively \\
    \hline
    z     & float & -    & Redshift \\
    \hline
    \end{tabular}%
  \label{tab:b}%
\end{table}%

	We have checked the SDSS images of the sources in Catalog A and B. Some of them show obvious characteristics of a binary system, so these quasars may be potential quasar pairs. The details of the two catalogs are given below.
	
	\subsection{Catalog A}
	\label{subsec:cataloga}
	The sources in Catalog A  have more than one matched source in $Gaia$ or SDSS within a $1^{\prime \prime}$ radius. They may be quasar pairs, star-quasar pairs, active galactic nuclei with obvious jets, or lensing objects. For the two sources with three $Gaia$ sources matched, the Simbad Astronomical Database \citep{wenger2000simbad} shows that there is a significant lensing effect near these two sources. Their SDSS IDs are 091127.61+055054.1 and 141546.24+112943.4, as mentioned above, they may be lensing objects, but more analysis is needed to determine that.
	
	To eliminate the interference of foreground stars, we mark some 5-parameter and 6-parameter sources with significant parallaxes and proper motions, which means that they might be star-quasar pairs. If at least one source in a pair has $|\varpi/\sigma_{\varpi}| > 5$, or $|\mu_{\alpha*}/\sigma_{\mu_{\alpha*}}| > 5$, or $|\mu_{\delta}/\sigma_{\mu_{\delta}}| > 5$, the pair is marked as star-quasar pair. According to this criterion, 62 pairs are preliminarily identified as star-quasar pairs.
	
\begin{figure}[!htbp]
	\centering
	\begin{tabular}{c}
		\includegraphics[width=0.18\textwidth]{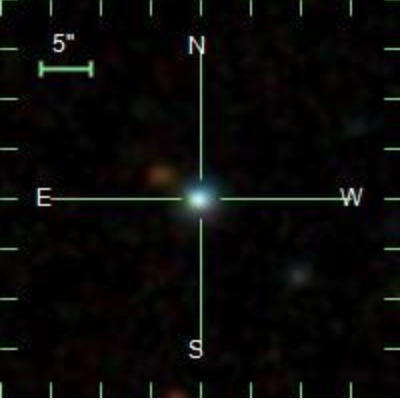}
		\small(A)
	\end{tabular}
	\begin{tabular}{c}
		\includegraphics[width=0.18\textwidth]{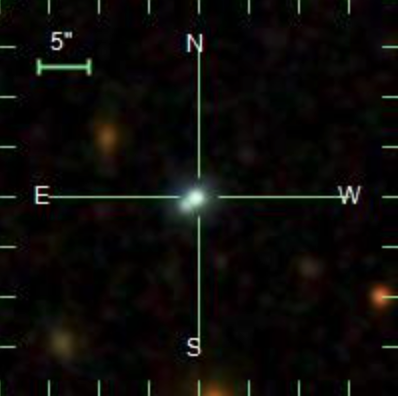}
		\small(B)
	\end{tabular}
	\begin{tabular}{c}
		\includegraphics[width=0.18\textwidth]{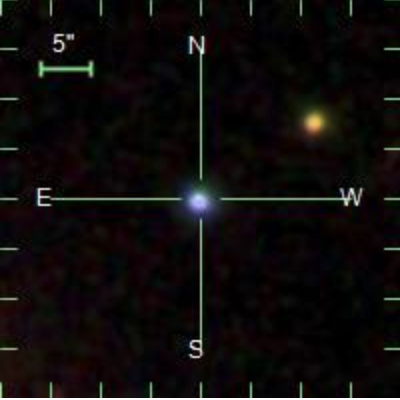}
		\small(C)
	\end{tabular}
	\begin{tabular}{c}
		\includegraphics[width=0.18\textwidth]{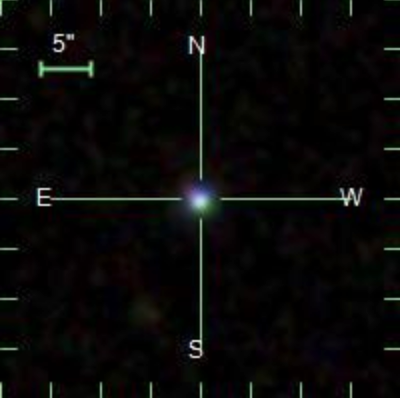}
		\small(D)
	\end{tabular}
	
	\begin{tabular}{c}
		\includegraphics[width=0.18\textwidth]{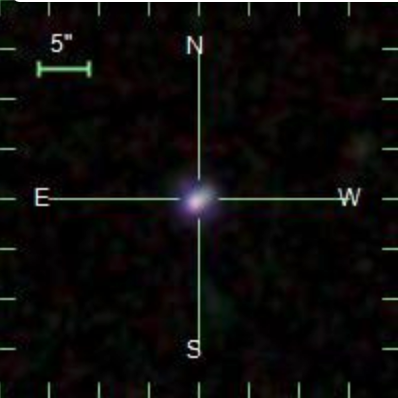}
		\small(E)
	\end{tabular}
	\begin{tabular}{c}
		\includegraphics[width=0.18\textwidth]{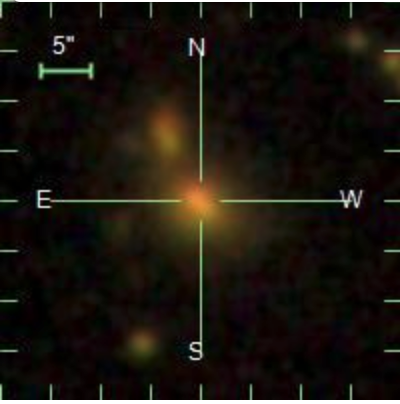}
		\small(F)
	\end{tabular}
	\begin{tabular}{c}
		\includegraphics[width=0.18\textwidth]{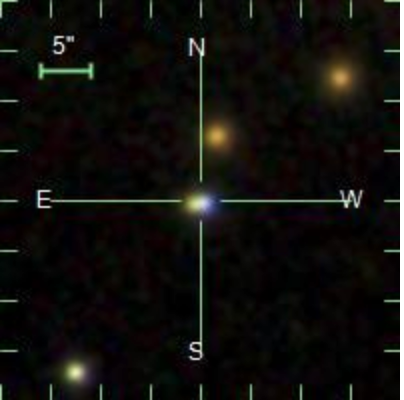}
		\small(G)
	\end{tabular}
	\begin{tabular}{c}
		\includegraphics[width=0.18\textwidth]{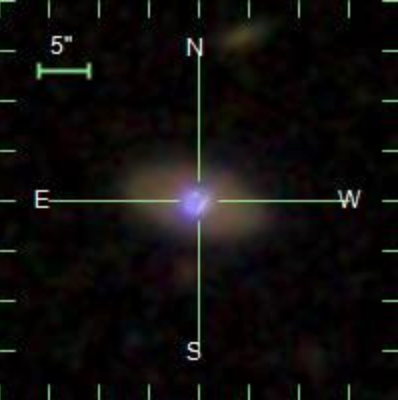}
		\small(H)
	\end{tabular}
	\caption
	{\small Eight SDSS images of sources in Catalog A, the top panels (A, B, C, D) are four point-like sources, while the bottom panels (E, F, G, H) are four extended sources.}
	\label{fig:cataloga}%
\end{figure}

	There are 64 extended sources and 91 point-like sources contained in Catalog A. Fig. \ref{fig:cataloga} shows several bright sources in Catalog A. For the point-like sources, most of them only have one optical center except Fig. \ref{fig:cataloga} (B), but the $Gaia$ high-precision observation indicates that there is more than one source in $1^{\prime \prime}$ radius of each SDSS position. Therefore more observations are needed for identifying if they are quasar pairs. For the extended sources, some of them exhibit obvious galaxy structures, such as Fig. \ref{fig:cataloga} (F), (H), while other extended sources may be caused by bright jets. In addition, the mean redshift of the extended sources is 1.19, and the average is 1.69 for the point-like sources. Therefore, these point-like sources are very important for studying high-redshift quasar pairs.

	\subsection{Catalog B}
	
	The sources in Catalog B are abnormal quasars, whose astrometric observation parameters deviate significantly from the entire sample. In $Gaia$ EDR3, all kinds of sources must be solitary. It means if there are multiple sources found within a $0.18^{\prime \prime}$ radius, the database will only keep one source with a flag named ``duplicated\_source''  \citep{lindegren2021gaia}. Although this flag does not definitely indicate that the source is a binary, it can be used as a reference to assess the reliability of the catalog. The proportion of duplicate sources in Catalog B is 11/44 (25\%), while the ratios in SDSS DR16Q and $Gaia$ FRS are 0.9\% and 0.6\%, respectively, which shows that our selection criteria are effective.
	
\begin{figure}[!htbp]
		\centering
		\begin{tabular}{c}
			\includegraphics[width=0.18\textwidth]{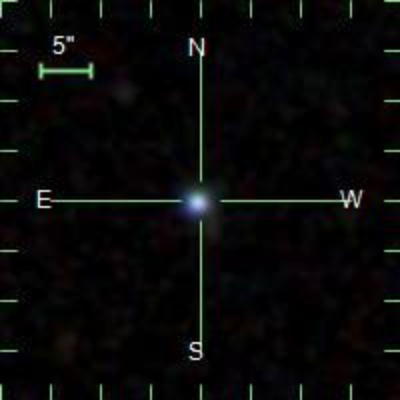}
			\small(A)
		\end{tabular}
		\begin{tabular}{c}
			\includegraphics[width=0.18\textwidth]{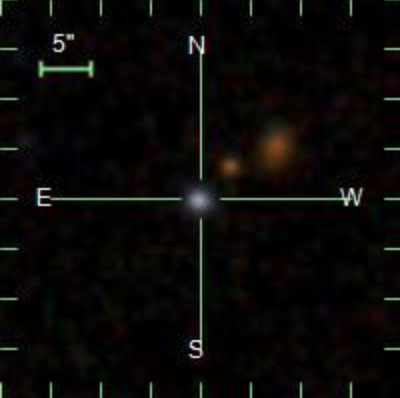}
			\small(B)
		\end{tabular}
		\begin{tabular}{c}
			\includegraphics[width=0.18\textwidth]{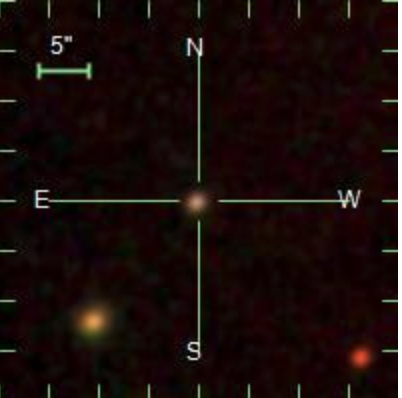}
			\small(C)
		\end{tabular}
		\begin{tabular}{c}
			\includegraphics[width=0.18\textwidth]{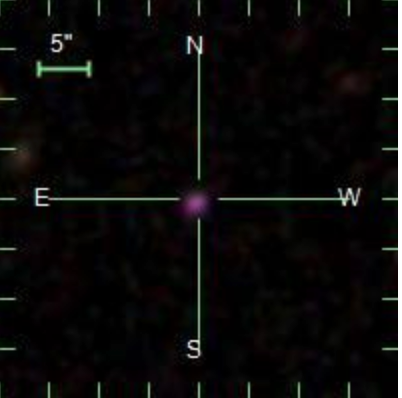}
			\small(D)
		\end{tabular}
		
		\begin{tabular}{c}
			\includegraphics[width=0.18\textwidth]{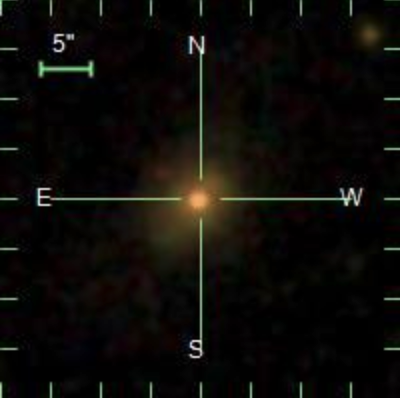}
			\small(E)
		\end{tabular}
		\begin{tabular}{c}
			\includegraphics[width=0.18\textwidth]{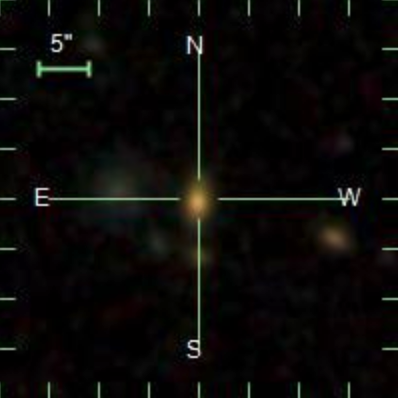}
			\small(F)
		\end{tabular}
		\begin{tabular}{c}
			\includegraphics[width=0.18\textwidth]{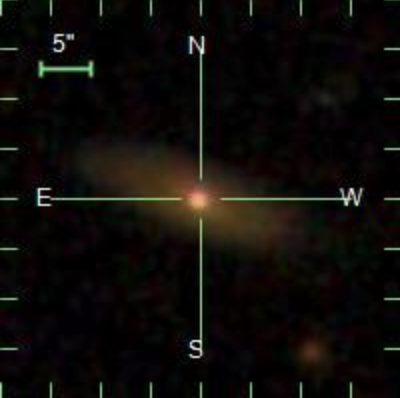}
			\small(G)
		\end{tabular}
		\begin{tabular}{c}
			\includegraphics[width=0.18\textwidth]{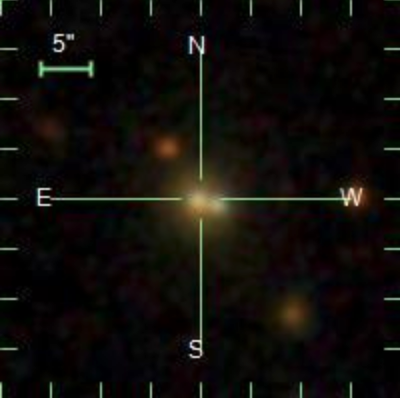}
			\small(H)
		\end{tabular}
		\caption
		{\small Eight SDSS images of sources in Catalog B, the top panels (A, B, C, D) are four point-like sources, while the bottom panels (E, F, G, H) are four extended sources.}
		\label{fig:catalogb}%
	\end{figure}

	Fig. \ref{fig:catalogb}, panel (A), (B), (E), (F) are four quasars with the flag ``duplicated\_source'', while the remaining four without this flag. Due to the low resolution of SDSS, there is no obvious difference between the images of duplicated and non-duplicated sources. Therefore, to further confirm whether these sources are quasar pairs or not, higher-resolution observations are needed, or maybe a method that combines spectral and light curves could be effective. Among the 25 extended sources, J115517.34+634622.0 is the only one with a redshift greater than 0.5, and its redshift is 2.9. The SDSS image of the source shown in Fig. \ref{fig:catalogb} (H) also exhibits a distinct dual optical center. Consistent with Catalog A, the average redshift of the point-like sources is 1.71, and the average redshift of the extended sources except J115517.34+634622.0 is 0.21. The huge redshift gap between the extended sources and the point sources may be because the extended structures of the long-distance high-redshift point sources are too faint to be observed.

	\section{Discussion}
	\label{sec:discussion}
	
	\subsection{Extened catalogs with different combination of the criteria}
	
	As we mentioned above, there are 0.9\% of SDSS DR16Q quasars with the flag ``duplicated\_source''. Although the proportion is very small, the number is huge. 4472 SDSS quasars with G $<$ 20.9 mag are duplicated sources, which indicates that Catalog B has poor completeness. In Eq \ref{criteria}, to improve the reliability of the catalog, we only select the sources that meet all the criteria. In fact,  each criterion can be used individually to select quasars with abnormal astrometric characteristics. 
	
\begin{figure}[!htbp]
	\centering
	\begin{tabular}{c}
		\includegraphics[width=0.43\textwidth]{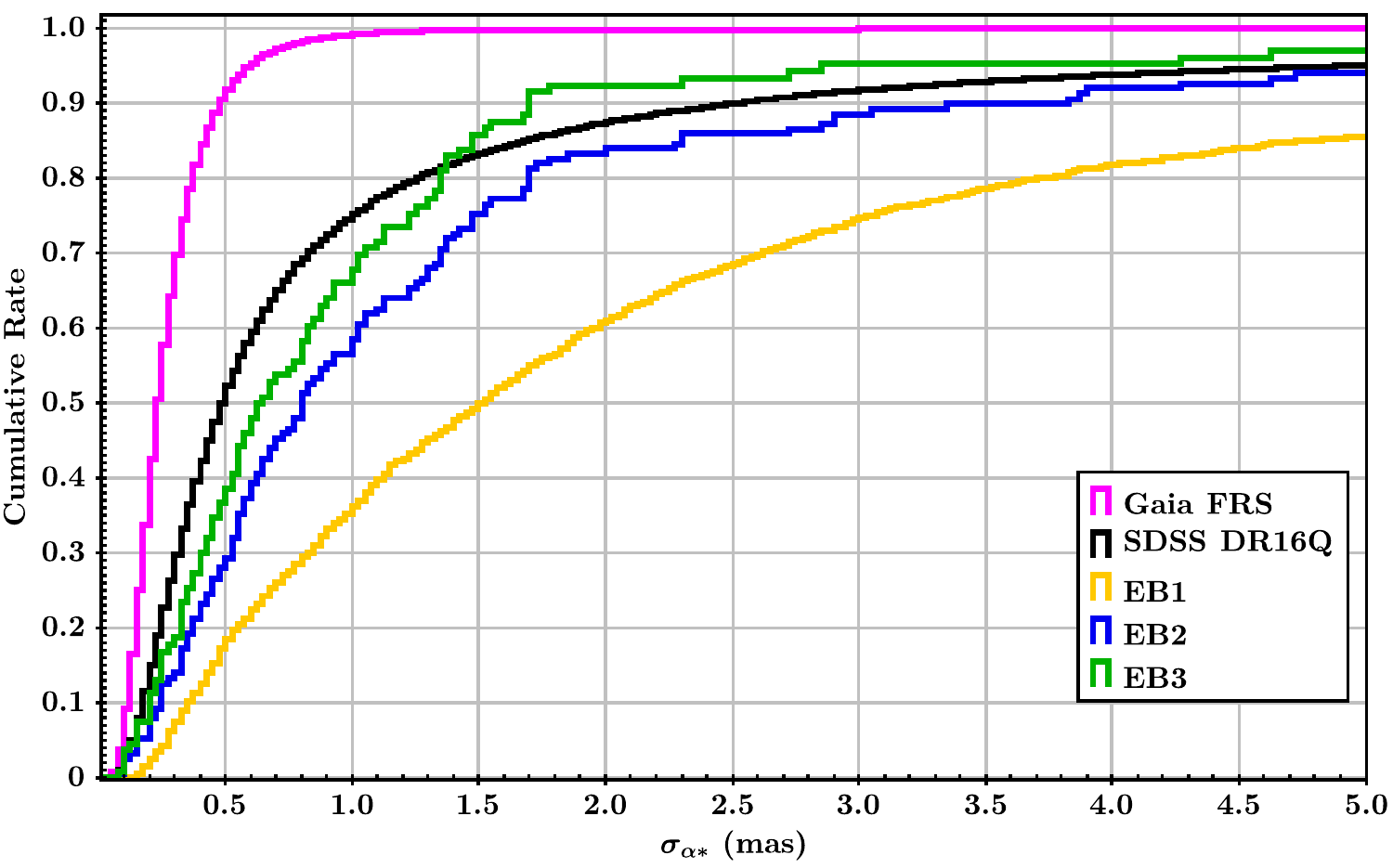}
		\small(A)
	\end{tabular}
	\begin{tabular}{c}
		\includegraphics[width=0.43\textwidth]{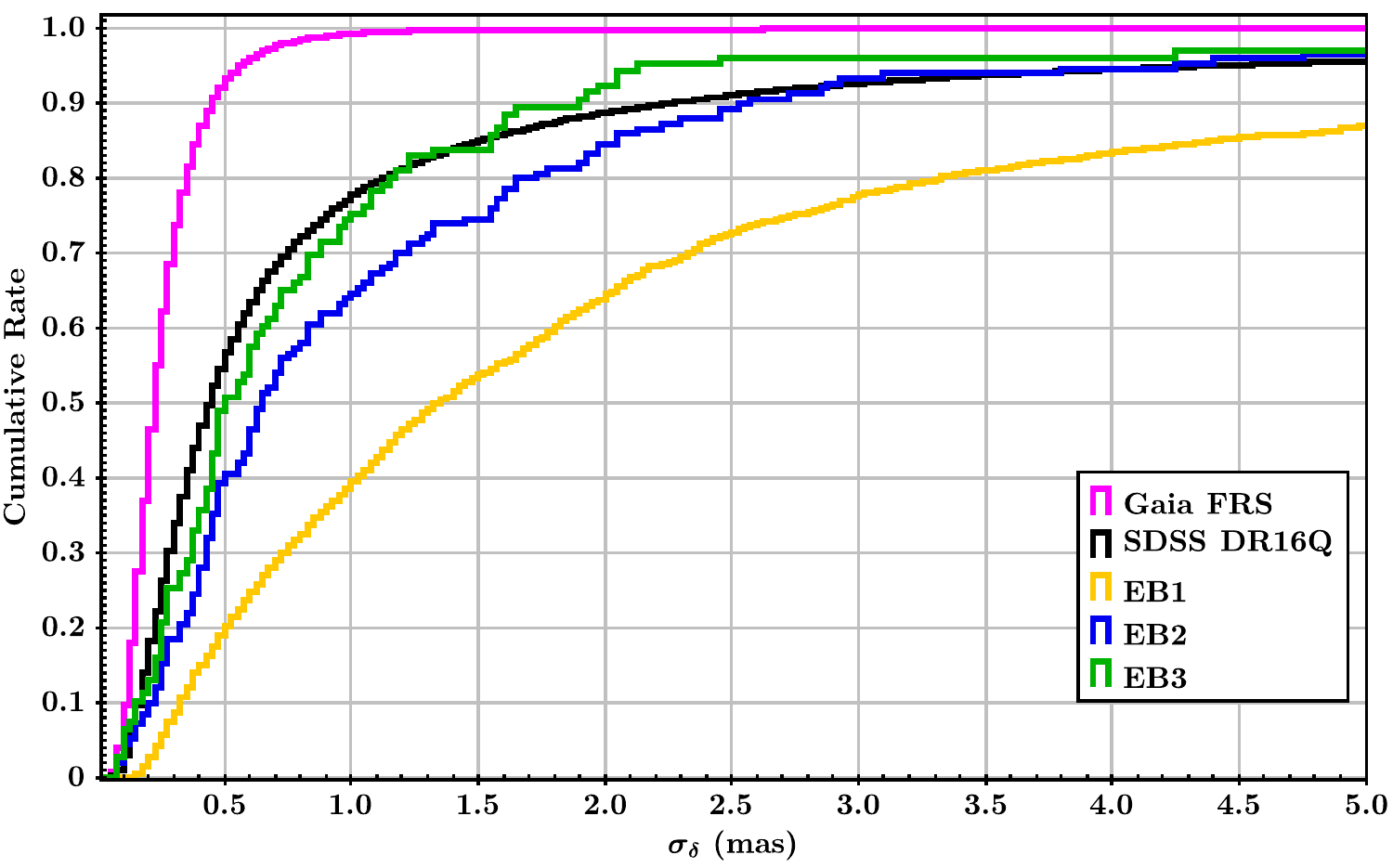}
		\small(B)
	\end{tabular}
	\caption
	{\small The cumulative distribution histogram for $\sigma_{\alpha*}$ (A) and $\sigma_{\delta}$ (B) of sources in $Gaia$ FRS, SDSS DR16Q, EB1, EB2 and EB3.}
	\label{fig:ebpositione}%
\end{figure}

	To select different kinds of abnormal sources, we consider three subsets of criteria in Eq \ref{criteria}: (1), the sources meet the criteria (i), (ii), (iii) and (vi); (2), the sources meet the criteria (iv), (v) and (vi); (3), the sources met the criteria (iv), (v) and (vi) but not the criteria (i), (ii) and (iii). The above three samples are respectively compiled into the Extended Catalog 1, 2, 3 of Catalog B (hereafter as EB1, EB2, EB3, respectively). According to their respective selection criteria, the sources in EB1 have bad fitting results in $Gaia$ EDR3, and the sources in EB2 may be visually resolved binaries.  EB3 contains the sources which have high percents of multi-peak but low noises, which means that there is another source near the EB3 source. Table \ref{tab:extendb} shows some statistical information of the three catalogs. Consistent with Catalog A and B, the catalogs with more extended sources have lower average redshift. There are hundreds of common sources in Extended catalogs of Catalog B and GEAC.  Fig. \ref{fig:ebpositione} shows that the sources in EB2 have slightly worse position precision than SDSS DR16Q sources, and the position precision of sources in EB1 is even worse than that of EB2. The number of duplicated sources in Table \ref{tab:extendb} also shows that we only select a small part of abnormal quasars.

\begin{table}[htbp]
  \centering
  \caption{Some details of Extended catalogs of Catalog B.}
    \begin{tabular}{|c|c|c|c|}
  \hline
          & EB1   & EB2   & EB3 \\
          \hline
Criteria & (i), (ii), (iii), (vi) & (iv), (v), (vi) & (iv), (v), (vi) not (i), (ii), (iii)\\
\hline
    Number & 1657  & 150   & 106 \\
\hline
    Number of extended sources & 648/1657 (39\%) & 52/150 (35\%) & 29/106 (27\%) \\
\hline
    Average redshift & 0.944 & 1.136 & 1.225 \\
\hline
   Number of common sources &\multirow{2} [0] {*} {771}&\multirow{2} [0] {*} {99}&\multirow{2} [0] {*} {83}\\
   in GEAC&  & & \\
\hline
   Number of common sources & \multirow{2} [0] {*} {36}&\multirow{2} [0] {*} {14}&\multirow{2} [0] {*} {14}\\
   in FRS& & &\\
\hline
Number of duplicated sources& 79 & 12  & 1 \\
\hline

    \end{tabular}%
  \label{tab:extendb}%
\end{table}%

\begin{figure}[h!]
	\begin{center}
		\includegraphics[width=0.7\textwidth]{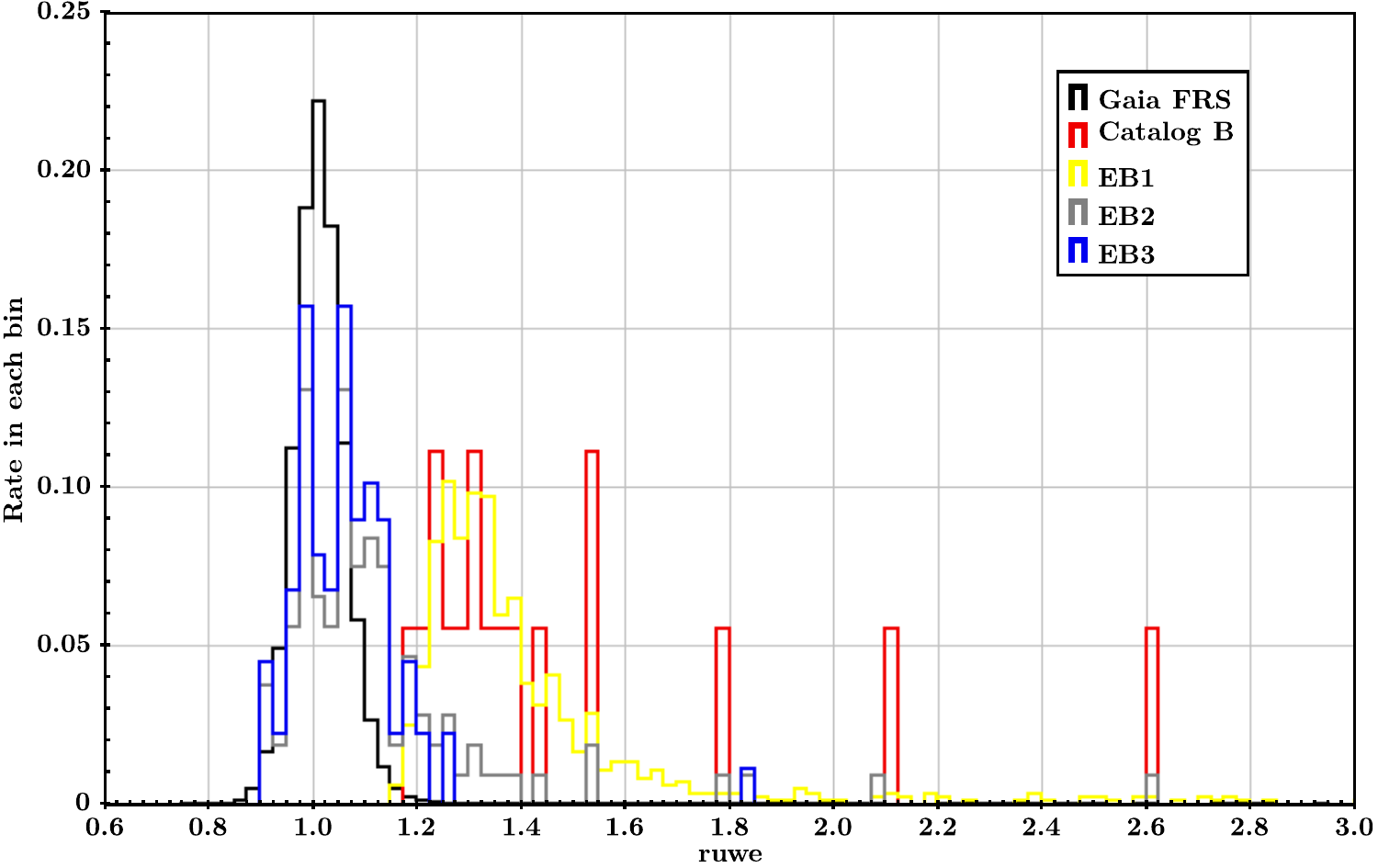}
	\end{center}
	\caption{Distribution histogram of the ruwe of sources in $Gaia$ FRS , Catalog B , EB1 , EB2 and EB3.}
	\label{fig:ruwe}
\end{figure}

	In addition to the above criteria, the renormalized unit weight error (ruwe) may also be a criterion that can be used to select binaries. In $Gaia$ Data Release 2 (DR2), ruwe $>$ 1.4 indicates that the source is a non-single star, however, this value is set to null for the 2-parameter sources in EDR3. As \citet{lindegren2021gaia} emphasized, both the ruwe and excess source noise quantify the disagreement between the $Gaia$ observations and the best-fitting model. Fig. \ref{fig:ruwe} shows that the ruwe of the sources in Catalog B is greater than that of FRS sources, which means that our criteria selecting binaries are effective. The ruwe of sources in EB1 is significantly higher than that of other samples, which proves that excess source noise and ruwe are consistent with each other. Therefore, ruwe is also a reliable indicator that can be used to select binary objects and may play an important role in our future releases.

	Apart from the SDSS DR16Q, there are many other reliable quasar catalogs such as the Large Bright Quasar Survey  \citep{hewett1995large}, the INT Wide Angle Survey  \citep{sharp2001first}, and the quasars from Large Sky Area Multi-Object Fiber Spectroscopic Telescope (LAMOST)  \citep{zhao2012lamost}. With our method of selecting abnormal astrometric quasars, a large number of quasar pair candidates will be selected.

	\subsection{Identification of Quasar Pairs and Lensed Quasars}

	In section \ref{sec:Data and Selection}, we have described the selection criteria and hence obtained two samples of abnormal quasars denoted as Catalog A and B. It is also interesting to explore the nature of these sources, whether they are quasar pairs, lensing images or containing jet-like structures. With the high-resolution observations from Hubble Space Telescope (HST), we could firstly resolve the general structures of these abnormal quasars, and further analysis of the corresponding spectra and light curves will be needed for a detailed classification.
	
\begin{figure}[h!]
		\begin{center}
			\includegraphics[width=0.7\textwidth]{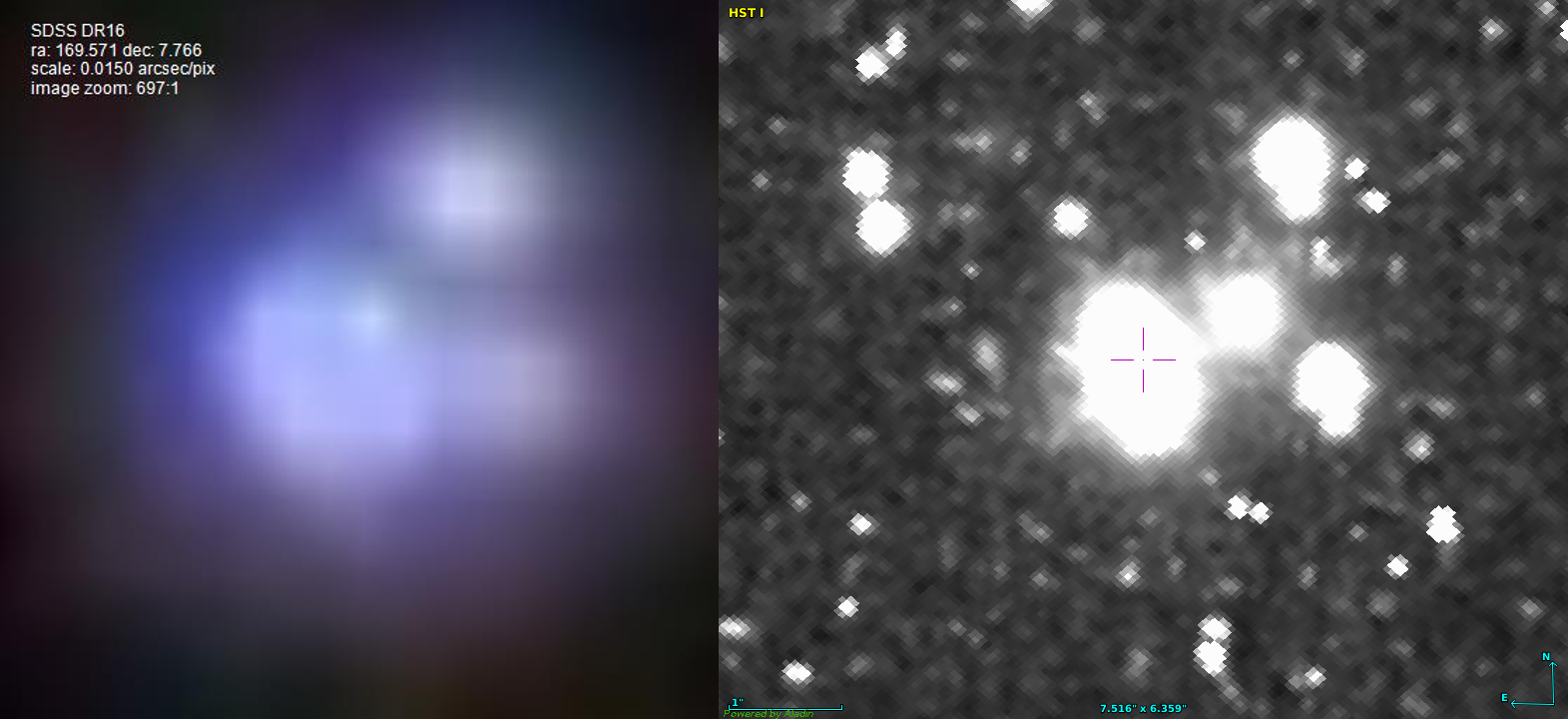}
		\end{center}
		\caption{SDSS (left side) and HST optical image (right side) of the lensed quasar SDSS 111816.94+074558.2, respectively. For the SDSS image, the resolution is 0.015 arcsec per pixel with 512 pixel in total. While for the HST image, the FOV is 7.516" $\times$ 6.359" and the scale 1" is labelled as the light blue line segment in the right panel.}
		\label{fig:pape2}
	\end{figure}

	There are about a dozen sources in Catalog A and Catalog B that have been identified as lensed quasars in the literatures. For example, Fig. \ref{fig:pape2} shows the SDSS image (left side) and the HST optical image (right side) of the lensed quasar SDSS 111816.94+074558.2 as identified in catalog A, already reported in  \citep{impey1998infrared,weymann1980triple}. We could see the advantage of the higher resolution of HST compared with SDSS when resolving the structure of those abnormal quasars. We also note that two similar pioneer works by \citet{shen2021hidden} and \citet{Chen_2022} have reported 2 and 43 AGN pair candidates, respectively, using the methods of varstrometry (i.e., excess astrometric noise). Among the sources we selected in this paper, 8 sources in Catalog A, 2 sources in Catalog B and 5 sources in EB1 have been reported as AGN pair candidates in their papers. However, both the target sample and the selection criteria are a little bit different. We will compare our results with them in future work.
	
	\section{Conclusion}
	\label{sec:conclusion}
	
	By cross-matching with other quasar catalogs, $Gaia$ EDR3 provides high-precision astrometric data for a large number of quasars, and a list of 1,614,173 quasar candidates are obtained, which could be used to establish the celestial reference frame in the optical band. However, during the selection process, many spectroscopically identified quasars showed abnormal astrometric characteristics, such as significant parallaxes and large proper motions. These quasars may come with astrometric jitter detectable with $Gaia$ data.  Therefore, with several $Gaia$ parameters describing the goodness of data fitting, quasars with abnormal astrometric characteristics could be selected. The selected quasars can form a group of quasar pair candidates.
	
	We propose a series of criteria for selecting abnormal quasars based on $Gaia$ astrometric data. Since $Gaia$ EDR3 contains 344 million 2-parameter sources, this means that these sources have only positional parameters. Our criteria do not rely on the complete data of parallax and proper motion, but depend on the goodness of fit to the observed data. With these criteria, two catalogs are obtained. Catalog A contains 155 SDSS quasars with more than one $Gaia$ matched within a $1^{\prime \prime}$ radius. Catalog B contains 44 SDSS quasars whose $Gaia$ observations are significantly different from the best-fitting standard astrometric model. The percentages of extended sources in Catalogs A and B are 41.3\% and 56.8\%, respectively. And in both catalogs, the mean redshift of the extended sources is significantly smaller than that of the point sources. 
	
	Although some of the SDSS images show obvious double star features, there are still many sources in our catalogs for which it is not possible to determine whether they are quasar pairs at the resolution of SDSS. Therefore, more high-resolution observations are needed to determine the fraction of quasar pairs of the catalogs in the future. In addition to SDSS DR16Q, many other quasar catalogs need to be further checked, so more efforts are needed to improve the selection criteria.

	There are 127 common sources between the GEAC quasars and our Catalog A and B, which should be excluded from GEAC for the purpose of establishing a reference frame. Besides, hundreds of common sources between Extended Catalogs of Catalog B and GEAC also show large position errors. The aspects of morphology and astrometric variability were crucial for selecting the quasars to form the reference frame \citep{ma2009second}. A perturbation in the disk of the host galaxy can cause a significant offset to the photocenter in the $Gaia$ observations \citep{popovic2012photocentric}. \citet{andrei2012morphology} used the morphological indexes in the $Gaia$ Initial QSO Catalog to indicate such influences. The host detection and characterization for about 1 million quasars will be released in the future release of  \href{https://www.cosmos.esa.int/web/gaia/dr3}{$Gaia$ DR3}. It might be interesting in the future to see if there is any correlation between the morphological parameters and the astrometric parameters mentioned in the current paper.

	\section*{Author Contributions}
	
   S. Liao is responsible for supervising the finding and selection of abnormal quasars from the $Gaia$ data. Q. Wu selected the $Gaia$ EDR3 and SDSS data with meticulous efforts and wrote the manuscript with help mainly from S. Liao and X. Ji. Besides, Z. Qi, Z. Zheng, Y. Zhang and T. An contributed to the physical interpretation and discussion. X. Ji and R. Lin contributed to collecting the SDSS data.

	\section*{Funding}
	This work has been supported by the Youth Innovation Promotion Association CAS, the grants from the Natural Science Foundation of Shanghai through grant 21ZR1474100, the National Natural Science Foundation of China (NSFC) through grants 12173069, 11703065, 11773051, 12022303 and the National Key R\&D Programme of China through grant 2018YFA0404603. We acknowledge the CAS Pioneer Hundred Talents Program and the science research grants from the China Manned Space Project with NO.CMS-CSST-2021-A12 and NO.CMS-CSST-2021-B10. 
	
	\section*{Acknowledgments}
	This work has made use of data from the European Space Agency (ESA) mission {\it $Gaia$} (\url{https://www.cosmos.esa.int/gaia}), processed by the {\it $Gaia$}
	Data Processing and Analysis Consortium (DPAC, \url{https://www.cosmos.esa.int/web/gaia/dpac/consortium}). Funding for the DPAC has been provided by national institutions, in particular the institutions participating in the {\it $Gaia$} Multilateral Agreement. We are very grateful for the data provided by SDSS. Funding for the Sloan Digital Sky Survey IV has been provided by the Alfred P. Sloan Foundation, the U.S. Department of Energy Office of Science, and the Participating Institutions. SDSS-IV acknowledges support and resources from the Center for High Performance Computing  at the University of Utah. The SDSS website is \href{www.sdss.org}{www.sdss.org}. This research has also made use of the SIMBAD database and the data of Hubble Space Telescope. We are also very grateful to the developers of the TOPCAT \citep{taylor2005topcat} software.

	\section*{Supplemental Data}
	The Supplemental Data of this paper contains Catalog A, Catalog B and three extended catalogs of Catalog B. The descriptions of Catalogs A and B are available in Table \ref{tab:a} and Table \ref{tab:b}, and for the three extended catalogs, we provide the $Gaia$ $source\_id$ and the corresponding $SDSS\_id$ of these sources.
	
	\section*{Data Availability Statement}
	The $Gaia$ EDR3 dataset is available in \href{https://gea.esac.esa.int/archive/}{$Gaia$ Archive}, and the SDSS data for this study can be found in \href{http://skyserver.sdss.org/dr16/en/home.aspx}{SDSS SkyServer DR16}.
	
	\bibliographystyle{frontiersinSCNS_ENG_HUMS} 
	\bibliography{test}
	
	
\end{document}